\documentclass[final,5p,times,twocolumn,sort&compress]{elsarticle}
\usepackage{graphicx}
\usepackage{mathptmx}      
\usepackage{amssymb}
\usepackage{amsfonts}
\usepackage[english]{babel}
\usepackage{amsmath}
\usepackage{enumerate}
\usepackage{epstopdf} 
\usepackage{multirow}
\usepackage{algorithm}
\usepackage{algorithmic}
\usepackage{circuitikz}
\usepackage{subcaption}
\usepackage[utf8]{inputenc}
\usepackage{algorithm}
\usepackage{algorithmic}
\usepackage{tikz}
\usepackage{tikz,graphicx,float}
\usetikzlibrary{shapes.geometric, arrows}
\usetikzlibrary{patterns}
\usetikzlibrary{automata,positioning}
\usetikzlibrary{calc,arrows,shapes,angles,quotes}
\usepackage{lipsum}
\usepackage{multicol}

\newtheorem{thm}{Theorem}[section]

\newtheorem{rem}[thm]{Remark}

\journal{}

\begin{document}
	
	\begin{frontmatter}
		
		
		
		\title{Bernstein-Vazirani Algorithm with A CCNOT-Based Oracle}
		
		\author{M.~H.~Annaby \corref{cor1}}
		\ead{mhannaby@cu.edu.eg}
		\address{Department of Mathematics, Faculty of Science, Cairo University, Giza 12613, Egypt}

		\begin{abstract}
		We introduce a quantum algorithm to solve Bernstein-Vazirani problem to recover secrete strings, using quantum oracles that are based on the Toffoli (CCNOT) logic gate. As in the known algorithm, the proposed algorithm is a polynomial speed-up algorithm. Moreover, the proposed approach allows us to solve new problems.
		\end{abstract}
		
		
	\end{frontmatter}


\section{Introduction}
Let $n\ge1$ be a positive integer, and $\langle\Bbb V,+,\cdot\rangle$ be the vector space $\{0,1\}^n$ over the field $\Bbb F_2=\{0,1\}$. Elements of $\Bbb V$ are strings of length $n$ of zeros and ones of the form $|\underline{\bf x}\rangle=|x_0x_1\cdots x_{n-1}\rangle$. We also denote vectors  of $\Bbb V$ by $\underline{\bf x}$. Here, the addition $+$ of $\Bbb V$ is the bit-wise XOR  
\[
\begin{array}{lcl}
|\underline{\bf x}\rangle+|\underline{\bf y}\rangle&=&|x_0x_1\cdots x_{n-1}\rangle+|y_0y_1\cdots y_{n-1}\rangle\\
{}&=&|(x_0\oplus y_0)\cdots (x_{n-1}\oplus y_{n-1})\rangle.
\end{array}
\]
Thus, the additive identity is $|\underline {\bf {0}}\rangle=|00\cdots0\rangle$ and every member of $\Bbb V$ is its own additive inverse. The scalar multiplication is
\[
\begin{array}{lcl}
0\cdot|\underline{\bf x}\rangle&=&0\cdot|x_0x_1\cdots x_{n-1}\rangle\\
{}&=&|(0\land x_0)(0\land x_1)\cdots (0\land x_{n-1})\rangle=|\underline {\bf {0}}\rangle,
\end{array}
\]
and
\[
\begin{array}{lcl}
	1\cdot|\underline{\bf x}\rangle&=&1\cdot|x_0x_1\cdots x_{n-1}\rangle\\
	{}&=&|(1\land x_0)(1\land x_1)\cdots (1\land x_{n-1})\rangle=|\underline {\bf {x}}\rangle,
\end{array}
\]
Moreover $\Bbb V$ is an Euclidean space with the inner product 
\[
\begin{array}{lcl}
	|\underline{\bf x}\rangle\odot|\underline{\bf y}\rangle&=&|x_0x_1\cdots x_{n-1}\rangle\odot|y_0y_1\cdots y_{n-1}\rangle\\
	{}&=&(x_0\land y_0)\oplus(x_1\land y_1)\oplus\cdots\oplus (x_{n-1}\land y_{n-1}).
\end{array}
\]
Hereafter, $\lor, \land, \neg, \oplus$ respectively denote the inclusive OR, the AND, the NOT and the exclusive OR (XOR) logic gates. Therefore, $\Bbb V$ is $n$-dimensional with the orthogonal basis 
\begin{equation}\label{eq01}
|\underline{\bf e_0}\rangle=|0\cdots01\rangle,\,|\underline{\bf e_1}\rangle=|0\cdots010\rangle,\;\cdots|\underline{\bf e_{n-1}}\rangle=|10\cdots00\rangle,
\end{equation}
cf. eg. \cite[Chapter 4]{LX1} for details.

We also denote the bases states of $\Bbb C^{\otimes n}$, by $|\underline{\bf {0}} \rangle,|\underline {\bf {1}}\rangle,\cdots |\underline{\bf {2^{n-1}}}\rangle$, where
\[
|\underline {\bf {0}}\rangle=|00\cdots0\rangle,\;|\underline {\bf {1}}\rangle=|00\cdots01\rangle,\;\cdots, |\underline {\bf {2^{n-1}}}\rangle=|11\cdots1\rangle.
\]

Let $f:\Bbb V\to \{0,1\}$ be a given function, such that  there exists a secret key $|\underline{\bf \varGamma}\rangle$, for which $f(\underline{\bf x})=|\underline{\bf x}\rangle\odot|\underline{\bf \varGamma}\rangle$. Bernstein-Vazirani problem (BVP) is to find ${|\underline{\bf \varGamma}}\rangle$. A classical solution can be achieved in at most $n$ computations of $f$ on the basis (\ref{eq01}). In such a situation, if  ${|\underline{\bf\varGamma}}\rangle=|\gamma_0 \gamma_1\cdots \gamma_{n-1}\rangle$, then $\gamma_j=f(\underline{\bf e_j}), \, j=0,\cdots,n-1$.

Bernstein-Vazirani algorithm (BVA), cf. \cite{BV011}, solves BVP in a single quantum measurement by applying  the circuit 
\begin{equation}
	\label{eq02}
	\left(H^{\otimes n}\otimes I\right) U_f \left(H^{\otimes n}\otimes H\right) |\underline{\bf {0}}1\rangle,
\end{equation}
where $H$ is the Hadamard-Walsh gate, $I$ is the identity and $U_f$ is the quantum oracle
\begin{equation}
	\label{eq03}
	U_f|\underline{\bf {x}}y\rangle=|\underline{\bf {x}},f(\underline{\bf {x}})\oplus y\rangle.
\end{equation}
Then BVA carries out a single measurement to the top qubit, which will be 100\% $|\underline{\bf{\varGamma}}\rangle$.

It is noted that the circuit (\ref{eq02}) is the circuit associated with Deutsch-Jozsa's algorithm, \cite{DJA1} with the same oracle (\ref{eq02}). Therefore, the oracle $U_f$ plays the main role in both algorithms. It is our aim to introduce another quantum oracle to solve BVP. We define two types of oracles, which are  based on the Toffolli (CCNOT) logic gate to solve BVP and other problems as well. Thus, an additional qubit is added to the quantum circuit, which produces entanglements and a straightforward application of $H^{\otimes n}$ will not produce the desired measurable state. However, if a change of phase oracle is also added, the algorithm will work, giving the correct result in a single measurement, which preserves the polynomial speed-up advantage of the quantum computing over the classical one.

Because of the produced entanglements, this approach will not only solve BVP with different oracles, but it allows us to solve new problems as well. This is an example of the power of entanglement, \cite{JL01}. For example, we give an algorithm to solve the following problem, which we denote by PI.

\noindent{\bf Problem PI.} Let $f:\Bbb V\to \{0,1\}$ be a given function, such that  there exists a secret key $|\underline{\bf \varGamma}\rangle$, for which $f(\underline{\bf x})=(|\underline{\bf x}\rangle+|\underline{\bf \varGamma}\rangle)\odot|\underline{\bf \varGamma}\rangle$. Find $|\underline{\bf \varGamma}\rangle$

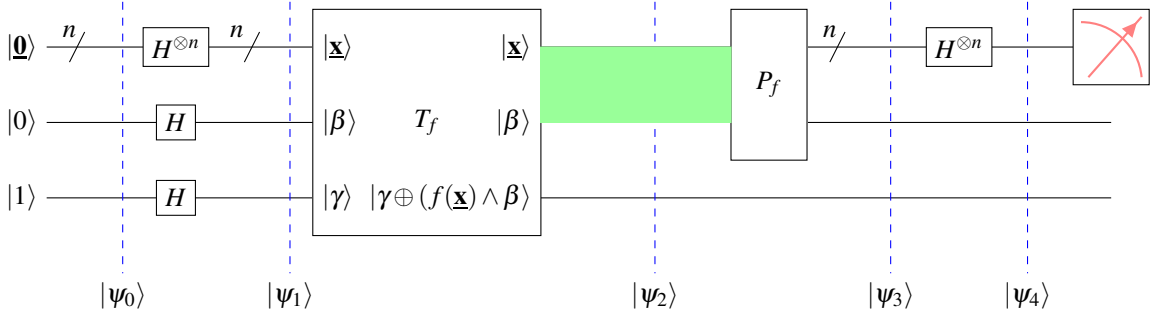
\begin{figure*}
	\centering
	\begin{tikzpicture}[y=-1cm,scale=2.00]
		\foreach\i in {0,1}
		\draw (7,\i) -- (0,\i);
		\draw[fill=white] (1.75,-0.25) rectangle node {$T_f$} (3.25,1.25);
		\draw[dashed,blue] (0.5,-0.3) -- (0.5,1.5) node[below,black] {$\lvert\psi_0\rangle$};
		\draw[dashed,blue] (1.6,-0.3) -- (1.6,1.5) node[below,black] {$\lvert\psi_1\rangle$};
		\draw[dashed,blue] (4.00,-0.3) -- (4.00,1.5) node[below,black] {$\lvert\psi_2\rangle$};
		\draw[dashed,blue] (5.55,-0.3) -- (5.55,1.5) node[below,black] {$\lvert\psi_3\rangle$};
		\draw[dashed,blue] (6.45,-0.3) -- (6.45,1.5) node[below,black] {$\lvert\psi_4\rangle$};
		\node[right] at (1.75,0) {$|\underline{\bf {x}}\rangle$};
		\node[left]  at (3.25,0) {$|\underline{\bf {x}}\rangle$};
		\node[right] at (1.75,0.5) {$|\beta\rangle$};
		\node[left]  at (3.25,0.5) {$|\beta\rangle$};
		\node[right] at (1.75,1) {\strut$|\gamma\rangle$};
		\node[left]  at (3.25,1) {\strut$\lvert \gamma\oplus(f(\underline{\bf {x}})\land \beta\rangle$};
		
		\draw (7,0.5)--(3.25,0.5);
		\draw (1.75,0.5)--(0,0.5) node[left] {$|0\rangle$};
		\node[left] at (0,0){$|\underline{\bf {0}}\rangle$};
		\node[left] at (0,1) {$|1\rangle$};
		\draw (0.85,0.5) node[draw,fill=white] {$H$};
		\draw (0.85,0.0) node[draw,fill=white] {$H^{\otimes n}$};
		\draw (6.00,0.0) node[draw,fill=white] {$H^{\otimes n}$};
		\draw (0.85,1.0) node[draw,fill=white] {$H$};
		\draw[fill=white] (4.5,-0.25) rectangle node {$P_f$} (5.00,0.75);
		\draw[fill=white] (6.75,-0.25) rectangle node {} (7.25,0.25);
		\draw[thick,red!50] (7.2,0.2) arc (355:270:0.4);
		\draw [thick,red!50] (6.85,0.2) -- (7.2,-0.2) node[red!50,
		currarrow,
		pos=0.8, 
		xscale=1,
		sloped,
		scale=1] {};
		\draw[] (0.25,-0.1)--(0.15,0.1);
		\node [above] at (0.15,0.0) {$n$};
		\draw[] (1.4,-0.1)--(1.3,0.1);
		\node [above] at (1.25,0.0) {$n$};
		\draw[] (5.25,-0.1)--(5.15,0.1);
		\node [above] at (5.15,0.0) {$n$};
		\draw[fill,green!40,opacity=0.5] (3.25,0.0) rectangle  (4.50,0.5);
	\end{tikzpicture}
	\caption{The circuit of the proposed Algorithm to solve BVP.}
	\label{Fig1}
\end{figure*}


\section{BV-Type Algorithms}


The first step to derive the proposed algorithm is to define the oracles. As is mentioned above two different oracles will be implemented to solve BVP in a single measurement. The first oracle depends on the Tofolli, CCNOT gate $T$, for which when it is applied to a 3-bit classical register $|\alpha\beta\gamma\rangle$, a classical 3-bit register $|\alpha\beta,\gamma\oplus(\beta\land\gamma)\rangle$ is produced. Toffoli gate is universal and its matrix is unitary and Hermitian, \cite{YM1}. Therefore, the first oracle is defined to be 
\begin{equation}
	\label{eq04}
	T_f|\underline{\bf {x}}\beta\gamma\rangle=|\underline{\bf {x}}\beta,\gamma\oplus(f(\underline{\bf {x}})\land\beta)\rangle. 
\end{equation}
As we have mentioned above, the algorithm requires the application of a change of phase oracle.  It is defined to be:

\begin{equation}
	\label{eq04}
	P_f|\underline{\bf {x}}\gamma\rangle= e^{-i\pi(f(\underline{\bf {x}})\land\neg\gamma)}|\underline{\bf {x}}\gamma\rangle. 
\end{equation}
The matrices representing the oracles $T_f$ and $P_f$ are unitary. Thus, unitarity of quantum operators are preserved. In addition, both matrices are Hermitian. For $P_f$, this is clear as it is diagonal, with diagonal entries $\pm1$.  The matrix of $T_f$ is a $2^{n+2}\times 2^{n+2}$ matrix with entries, which are zeros or ones and each row has exactly one 1 and the rest are zeros. So, to see that the matrix of $T_f$ is unitary, it suffices to prove that for all $f$ and all $|\underline{\bf {x}}\rangle$ of $\Bbb V$, the states
\begin{equation}
	\label{eq09}
T_f|\underline{\bf {x}}00\rangle,\;T_f|\underline{\bf {x}}01\rangle,\;T_f|\underline{\bf {x}}10\rangle,\;T_f|\underline{\bf {x}}11\rangle
\end{equation}
are different bases vectors. Indeed, let $\underline{\bf {x}}\in\Bbb V,$ be arbitrary. If $f(\underline{\bf {x}})=0,$ then 

\[T_f|\underline{\bf {x}}\beta\gamma\rangle=|\underline{\bf {x}}\beta,\gamma\oplus(0\land\beta)\rangle=|\underline{\bf {x}}\beta\gamma\rangle,
\]
i.e. (\ref{eq09}) are different bases states. Similarly, if $f(\underline{\bf {x}})=1,$ then also 
\[
\begin{array}{lcl}
T_f|\underline{\bf {x}}00\rangle&=&|\underline{\bf {x}}0,0\oplus(1\land0)\rangle=|\underline{\bf {x}}00\rangle,\\
T_f|\underline{\bf {x}}01\rangle&=&|\underline{\bf {x}}0,1\oplus(1\land0)\rangle=|\underline{\bf {x}}01\rangle\\
T_f|\underline{\bf {x}}10\rangle&=&|\underline{\bf {x}}1,0\oplus(1\land1)\rangle=|\underline{\bf {x}}11\rangle\\
T_f|\underline{\bf {x}}11\rangle&=&|\underline{\bf {x}}1,1\oplus(1\land1)\rangle=|\underline{\bf {x}}10\rangle.
\end{array}
\]
This does not only prove unitarity of $T_f$, but also it proves self-adjointness since the states $|\underline{\bf {x}}10\rangle$ and $|\underline{\bf {x}}11\rangle$ interchange the ones in their symmetric places.

\vskip0.1in

The proposed algorithm to solve BVP is:

\vskip0.1in

\noindent{\bf Step I.} Run the quantum circuit
\begin{equation}
\label{eq05}
(H^{\otimes n}\otimes I\otimes I)\, (P_f\otimes I)\, T_f \, (H^{\otimes n}\otimes H\,\otimes H)|\underline{\bf {0}}01\rangle.
\end{equation}

\noindent{\bf Step II.} Measure the top state, which will be certainly $|\underline{\bf \varGamma}\rangle$.

The circuit of the above algorithm is depicted in Figure 1 and it evolves as follows: $|\psi_0\rangle=|\underline{\bf {0}}01\rangle,$ 
\[
|\psi_1\rangle=\displaystyle\frac{1}{\sqrt{2^n}}\sum_{x=0}^{2^n-1} |\underline{\bf {x}}\rangle\, \otimes |+\rangle\,\otimes |-\rangle,
\]
where $|+\rangle,|-\rangle$ are the spin-to-left and the spin-to-right states
\[
|+\rangle=\displaystyle\frac{|0\rangle+|1\rangle}{\sqrt2},\quad
|-\rangle=\displaystyle\frac{|0\rangle-|1\rangle}{\sqrt2}.
\]
To compute $|\psi_2\rangle,$ we notice that
\[
\begin{array}{lcl}
	T_f|\underline{\bf {x}}00\rangle&=&|\underline{\bf {x}}0,0\oplus(f(\underline{\bf {x}})\land 0)\rangle=|\underline{\bf {x}}0\rangle\,|0\rangle,\\
		T_f|\underline{\bf {x}}01\rangle&=&|\underline{\bf {x}}0,1\oplus(f(\underline{\bf {x}})\land 0)\rangle=|\underline{\bf {x}}0,\rangle\,|1\rangle.
\end{array}
\]
Thus
\[
T_f|\underline{\bf {x}}0\rangle\,|-\rangle=|\underline{\bf {x}}0\rangle\,|-\rangle.
\]
But
\[
\begin{array}{lcl}
T_f|\underline{\bf {x}}10\rangle&=&|\underline{\bf {x}}1,0\oplus(f(\underline{\bf {x}})\land 1)\rangle=|\underline{\bf {x}}1 \rangle|f(\underline{\bf {x}}) \rangle,\\
T_f|\underline{\bf {x}}11\rangle&=&|\underline{\bf {x}}1,1\oplus(f(\underline{\bf {x}})\land 1)\rangle=|\underline{\bf {x}}1 \rangle|\neg f(\underline{\bf {x}}) \rangle,
\end{array}
\]
i.e.
\[
T_f|\underline{\bf {x}}1\rangle\,|-\rangle=(-1)^{f(\underline{\bf {x}})}\,|\underline{\bf {x}}1\rangle\,|-\rangle.
\]
Therefore, 
\[
|\psi_2\rangle=\displaystyle\frac{1}{\sqrt2\,\sqrt{2^n}}\sum_{x=0}^{2^n-1} \left[|\underline{\bf {x}}0\rangle+ (-1)^{f(\underline{\bf {x}})} |\underline{\bf {x}}1\rangle\right] \,\otimes |-\rangle.
\]

The application of the oracle $P_f$ changes the phase of each $|\underline{\bf {x}}0\rangle$ to be $(-1)^{f(\underline{\bf {x}})}|\underline{\bf {x}}0\rangle$. Therefore $|\psi_3\rangle=(P_f\otimes I)|\psi_2\rangle$ is
\[
\begin{array}{lcl}
|\psi_3\rangle&=&(P_f\otimes I)\left[\displaystyle\sum_{x=0}^{2^n-1}\displaystyle\frac{\left[|\underline{\bf {x}}0\rangle+ (-1)^{f(\underline{\bf {x}})} |\underline{\bf {x}}1\rangle\right]}{\sqrt2\,\sqrt{2^n}}  \,\otimes |-\rangle\right]\\
{}&=& \displaystyle\frac{1}{\sqrt2\,\sqrt{2^n}}\sum_{x=0}^{2^n-1} \left[(-1)^{f(\underline{\bf {x}})}|\underline{\bf {x}}0\rangle+ (-1)^{f(\underline{\bf {x}})} |\underline{\bf {x}}1\rangle\right] \,\otimes |-\rangle\\
{}&=& \displaystyle\frac{1}{\sqrt2\,\sqrt{2^n}}\sum_{x=0}^{2^n-1} (-1)^{f(\underline{\bf {x}})}\left[|\underline{\bf {x}}0\rangle+|\underline{\bf {x}}1\rangle\right]\otimes |-\rangle\\
{}&=&\displaystyle\frac{1}{\sqrt{2^n}}\left[\sum_{x=0}^{2^n-1}(-1)^{f(\underline{\bf {x}})}|\underline{\bf {x}}\rangle\right]\otimes |+\rangle\otimes |-\rangle.
\end{array} 
\]

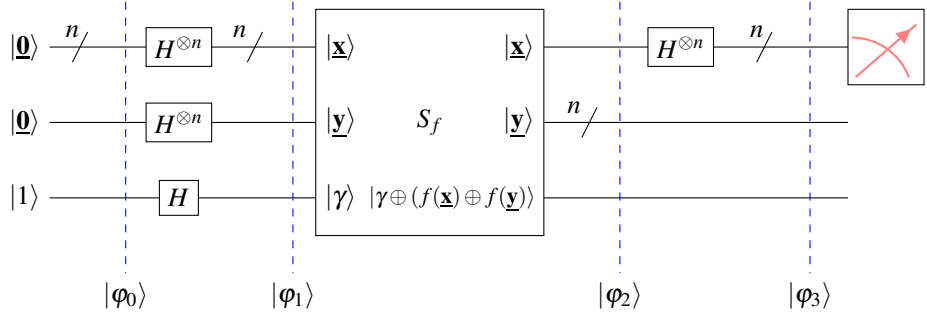
\begin{figure*}
	\centering
	\begin{tikzpicture}[y=-1cm,scale=2.00]
		\foreach\i in {0,1}
		\draw (5.25,\i) -- (0,\i);
		\draw[fill=white] (1.75,-0.25) rectangle node {$S_f$} (3.25,1.25);
		\draw[dashed,blue] (0.5,-0.3) -- (0.5,1.5) node[below,black] {$\lvert\varphi_0\rangle$};
		\draw[dashed,blue] (1.6,-0.3) -- (1.6,1.5) node[below,black] {$\lvert\varphi_1\rangle$};
		\draw[dashed,blue] (3.75,-0.3) -- (3.75,1.5) node[below,black] {$\lvert\varphi_2\rangle$};
		\draw[dashed,blue] (5.00,-0.3) -- (5.0,1.5) node[below,black] {$\lvert\varphi_3\rangle$};
		\node[right] at (1.75,0) {$|\underline{\bf {x}}\rangle$};
		\node[left]  at (3.25,0) {$|\underline{\bf {x}}\rangle$};
		\node[right] at (1.75,0.5) {$|\underline{\bf {y}}\rangle$};
		\node[left]  at (3.25,0.5) {$|\underline{\bf {y}}\rangle$};
		\node[right] at (1.75,1) {\strut$|\gamma\rangle$};
		\node[left,scale=0.85]  at (3.25,1) {\strut$\lvert \gamma\oplus(f(\underline{\bf {x}})\oplus f(\underline{\bf {y}})\rangle$};
		
		\draw (5.25,0.5)--(3.25,0.5);
		\draw (1.75,0.5)--(0,0.5) node[left] {$|\underline{\bf {0}}\rangle$};
		\node[left] at (0,0){$|\underline{\bf {0}}\rangle$};
		\node[left] at (0,1) {$|1\rangle$};
		\draw (0.85,0.5) node[draw,fill=white] {$H^{\otimes n}$};
		\draw (0.85,0.0) node[draw,fill=white] {$H^{\otimes n}$};
		\draw (0.85,1.0) node[draw,fill=white] {$H$};
		\draw (4.15,0.0) node[draw,fill=white] {$H^{\otimes n}$};
		\draw[fill=white] (5.25,-0.25) rectangle node {} (5.75,0.25);
		\draw[thick,red!50] (5.65,0.2) arc (340:270:0.4);
		\draw [thick,red!50] (5.30,0.2) -- (5.7,-0.15) node[red!50,
		currarrow,
		pos=0.8, 
		xscale=1,
		sloped,
		scale=1] {};
		\draw[] (0.25,-0.1)--(0.15,0.1);
		\node [above] at (0.15,0.0) {$n$};
		\draw[] (3.6,0.4)--(3.5,0.6);
		\node [above] at (3.45,0.5) {$n$};
		\draw[] (1.4,-0.1)--(1.3,0.1);
		\node [above] at (1.25,0.0) {$n$};
		\draw[] (4.75,-0.1)--(4.65,0.1);
		\node [above] at (4.65,0.0) {$n$};
	\end{tikzpicture}
	\caption{The circuit of the proposed Algorithm (\ref{eq07}) to solve the PI problem.}
	\label{Fig2}
\end{figure*}

But $(-1)^{f(\underline{\bf {x}})}=(-1)^{f(|\underline{\bf x}\rangle\odot|\underline{\bf \varGamma}\rangle)},$ and, cf. \cite{YM1}
\[
H^{\otimes n}|\underline{\bf \varGamma}\rangle=\displaystyle\frac{1}{\sqrt{2^n}}\left[\sum_{x=0}^{2^n-1}(-1)^{(|\underline{\bf x}\rangle\odot|\underline{\bf \varGamma}\rangle)}|\underline{\bf {x}}\rangle\right],
\]
i.e.
\[
|\psi_3\rangle=H^{\otimes n}|\underline{\bf \varGamma}\rangle\, \otimes |+\rangle\,\otimes |-\rangle.
\]
Since $H^{\otimes n}$ is its own inverse, then
\[
|\psi_4\rangle=|\underline{\bf \varGamma}\rangle\, \otimes |+\rangle\,\otimes |-\rangle.
\]
Measuring the top quantum register surely gives $|\underline{\bf \varGamma}\rangle$. 

\vskip0.1in  

\begin{rem}
It should be noted that in BVA {\textup{(\ref{eq02})}} the state $ U_f \left(H^{\otimes n}\otimes H\right) |\underline{\bf {0}}1\rangle$ has the form 
\[
U_f \left(H^{\otimes n}\otimes H\right) |\underline{\bf {0}}1\rangle=H^{\otimes n}|\underline{\bf \varGamma}\rangle\, \otimes |-\rangle,
\]
	while in the case of {\textup{(\ref{eq05})}}, the application of $P_f$ is necessary to separate the top quantum register. 
\end{rem}

\vskip0.1in

The algorithm to solve Problem PI also consists of two steps. It does not implement the oracles $T_f,\, P_f$, but another single oracle $S_f$, defined by
\[
S_f|\underline{\bf {x}}\,\underline{\bf {y}}\gamma\rangle={|\underline{\bf {x}}\,\underline{\bf {y}},\gamma\oplus(f(\underline{\bf {x}})\oplus f(\underline{\bf {y}})\rangle}.
\]
We carry out a similar analysis to that established above to show that the matrix of $S_f$ is unitary and Hermitian. Let $\underline{\bf {x}},\,\underline{\bf {y}}\in \Bbb V$ be arbitrary. If  $f(\underline{\bf {x}})=f(\underline{\bf {y}}),$ then $
S_f|\underline{\bf {x}}\,\underline{\bf {y}}\gamma\rangle={|\underline{\bf {x}}\,\underline{\bf {y}},\gamma\rangle},$ i.e. $|\underline{\bf {x}}\,\underline{\bf {y}}0\rangle$ and $|\underline{\bf {x}}\,\underline{\bf {y}}1\rangle$ are different state elements. Likewise, if $f(\underline{\bf {x}})\ne f(\underline{\bf {y}}),$ then $
S_f|\underline{\bf {x}}\,\underline{\bf {y}}\gamma\rangle={|\underline{\bf {x}}\,\underline{\bf {y}},\neg\gamma\rangle},$ i.e. also $|\underline{\bf {x}}\,\underline{\bf {y}}0\rangle$ and $|\underline{\bf {x}}\,\underline{\bf {y}}1\rangle$ are different state elements. Hence unitarity and self-adjointness are proved.

\vskip0.1in

The two-step algorithm proposed to solve problem PI is:

\vskip0.1in

\noindent{\bf Step I.} Run the quantum circuit
\begin{equation}
\label{eq07}
(H^{\otimes n}\otimes I\otimes I)\, S_f \, (H^{\otimes n}\otimes H^{\otimes n}\otimes H)|\underline{\bf {0}}\,\underline{\bf {0}}1\rangle.
\end{equation}

\noindent{\bf Step II.} Measure the top state, and it will be surely $|\underline{\bf \varGamma}\rangle$. 
\vskip0.1in

The evolution of this algorithm is exhibited in Figure \ref{Fig2}. To prove that the algorithm works, we compute its evolution. Indeed, $|\varphi_0\rangle=|\underline{\bf {0}}\,\underline{\bf {0}}1\rangle$. Now,
\[
\begin{array}{lcl}
|\varphi_1\rangle&=&\displaystyle\frac{1}{\sqrt{2^n}}\sum_{x=0}^{2^n-1} |\underline{\bf {x}}\rangle\, \otimes \displaystyle\frac{1}{\sqrt{2^n}}\sum_{y=0}^{2^n-1} |\underline{\bf {y}}\rangle\,\otimes |-\rangle\\
{}&=& \displaystyle\frac{1}{{2^n}}\sum_{y=0}^{2^n-1}\sum_{x=0}^{2^n-1} |\underline{\bf {x}}\rangle\,|\underline{\bf {y}}\rangle\, \otimes \, |-\rangle.
\end{array}
\]
The effect of $S_f$ on $|\varphi_1\rangle$ is different from $T_f$. For, 
\[
\begin{array}{lcl}
	S_f|\underline{\bf {x}}\,\underline{\bf {y}}0\rangle&=&|\underline{\bf {x}}\,\underline{\bf {y}},0\oplus(f(\underline{\bf {x}})\oplus f(\underline{\bf {y}}))\rangle=|\underline{\bf {x}}\underline{\bf {y}},f(\underline{\bf {x}})\oplus f(\underline{\bf {y}})\rangle,\\
	S_f|\underline{\bf {x}}\,\underline{\bf {y}}1\rangle&=&|\underline{\bf {x}}\,\underline{\bf {y}},1\oplus(f(\underline{\bf {x}})\oplus f(\underline{\bf {y}}))\rangle=|\underline{\bf {x}}\underline{\bf {y}},\neg(f(\underline{\bf {x}})\oplus f(\underline{\bf {y}}))\rangle.
\end{array}
\]
Therefore,
\[
S_f|\underline{\bf {x}}\,\underline{\bf {y}}\rangle\,|-\rangle=(-1)^{f(\underline{\bf {x}})\oplus f(\underline{\bf {y}})}|\underline{\bf {x}}\,\underline{\bf {y}}\rangle\,|-\rangle.
\]
Hence,
\begin{equation}\label{eq08}
|\varphi_2\rangle=\displaystyle\frac{1}{{2^n}}\sum_{y=0}^{2^n-1}\sum_{x=0}^{2^n-1} (-1)^{f(\underline{\bf {x}})\oplus f(\underline{\bf {y}})}|\underline{\bf {x}}\rangle\,|\underline{\bf {y}}\rangle\, \otimes \, |-\rangle.
\end{equation}
However, if $|\underline{\bf {x}}\rangle=|x_0\cdots x_{n-1}\rangle,\,|\underline{\bf {y}}\rangle=|y_0\cdots y_{n-1}\rangle,\,|\underline{\bf {\varGamma}}\rangle=|\gamma_0\cdots \gamma_{n-1}\rangle,$ then
\[
\begin{array}{lcl}
f(\underline{\bf {x}})&=&(\underline{\bf {x}}+\underline{\bf {\varGamma}})\odot \underline{\bf {\varGamma}}\\
{}&=&|(x_0\oplus\gamma_0)(x_1\oplus\gamma_1)\cdots(x_{n-1}\oplus\gamma_{n-1})\odot\underline{\bf {\varGamma}}\\
{}&=&((x_0\oplus\gamma_0)\land\gamma_0)\oplus\cdots\oplus ((x_{n-1}\oplus\gamma_{n-1})\land\gamma_{n-1})\\
{}&=&((x_0\land\gamma_0)\oplus\gamma_0)\oplus\cdots\oplus ((x_{n-1}\land\gamma_{n-1})\oplus\gamma_{n-1}).
\end{array}
\]
Similarly
\[
f(\underline{\bf {y}})=((y_0\land\gamma_0)\oplus\gamma_0)\oplus\cdots\oplus ((y_{n-1}\land\gamma_{n-1})\oplus\gamma_{n-1}).
\]
Therefore,
\[
\begin{array}{lcl}
f(\underline{\bf {x}})\oplus f(\underline{\bf {y}})&=&((x_0\land\gamma_0)\oplus(y_0\land\gamma_0))\oplus\cdots\\
{}&{}&\oplus((x_{n-1}\land\gamma_{n-1})\oplus(y_{n-1}\land\gamma_{n-1}))\\
{}&=&(|\underline{\bf {x}}\rangle\odot |\underline{\bf {\varGamma}}\rangle)\oplus (|\underline{\bf {y}}\rangle\odot |\underline{\bf {\varGamma}}\rangle).
\end{array}
\]
Substituting in (\ref{eq08}) yields
\[
\begin{array}{lcl}
|\varphi_2\rangle&=&\displaystyle\frac{1}{\sqrt{2^n}}\sum_{x=0}^{2^n-1} (-1)^{|\underline{\bf {x}}\rangle\odot |\underline{\bf {\varGamma}}\rangle} |\underline{\bf {x}}\rangle\\
{}&{}& \otimes \displaystyle\frac{1}{\sqrt{2^n}}\sum_{y=0}^{2^n-1} (-1)^{|\underline{\bf {y}}\rangle\odot |\underline{\bf {\varGamma}}\rangle} |\underline{\bf {y}}\rangle\,\otimes |-\rangle\\
{}&=& H^{\otimes n}|\underline{\bf {\varGamma}}\rangle\otimes H^{\otimes n}|\underline{\bf {\varGamma}}\rangle \otimes |-\rangle.
\end{array}
\]
Hence applying $H^{\otimes n}\otimes I\otimes I$ to $|\varphi_2\rangle$ and measuring the top state, we obtain $|\underline{\bf {\varGamma}}\rangle$, and the algorithm works. 
\begin{rem}
\mbox{}
\vspace{0.001cm}
\begin{enumerate}
	\item[$\bullet$] A classical solution of PI also has order $\mathcal{O}(n)$. To see this, Let $|\underline{\bf {k_j}}\rangle=\neg|\underline{\bf {e_j}}\rangle$, i.e. all entries of $|\underline{\bf {k_j}}\rangle$ are ones except for $0$ in the $j$-th position. Thus,
	\[
	|\underline{\bf {k_j}}\rangle=|1\cdots101\cdots1\rangle.
	\]
	Therefore, $f(\underline{\bf {k_j}})=\gamma_j\land \gamma_j=\gamma_j.$ Consequently, the proposed algorithm is a polynomial speed-up too.
	
	\item[$\bullet$] We can see that the BVA {\textup{(\ref{eq02})}} will not solve the problem PI. Indeed,  if we apply {\textup{(\ref{eq02})}} to $|\underline{\bf {0}}1\rangle$, we would get 
	\begin{equation*}
		\begin{array}{lcl}
			U_f \left(H^{\otimes n}\otimes H\right) |\underline{\bf {0}}1\rangle&=&\displaystyle\frac{1}{\sqrt{2^n}}\sum_{x=0}^{2^n-1} (-1)^{f(\underline{\bf {x}})}|\underline{\bf {x}}\rangle\otimes \, |-\rangle\\
			{}&=& \displaystyle\frac{1}{\sqrt{2^n}}\sum_{x=0}^{2^n-1} (-1)^{(|\underline{\bf {x}}\rangle+|\underline{\bf {\varGamma}}\rangle)\odot |\underline{\bf {\varGamma}}}|\underline{\bf {x}}\rangle\otimes \, |-\rangle.
		\end{array}
	\end{equation*}
	From the properties of the inner product, cf. {\textup{\cite{LX1}}}, 
	\[
	(|\underline{\bf {x}}\rangle+|\underline{\bf {\varGamma}}\rangle)\odot |\underline{\bf {\varGamma}}=(|\underline{\bf {x}}\rangle\odot|\underline{\bf {\varGamma}}\rangle)\oplus (|\underline{\bf {\varGamma}}\rangle\odot|\underline{\bf {\varGamma}}\rangle).
	\]
	I.e. the application of $H^{\otimes n}\otimes I$ and the measurement of the top states reveals nothing unless $|\underline{\bf {\varGamma}}\rangle\odot|\underline{\bf {\varGamma}}\rangle=0$.
	
\item[$\bullet$] Instead of measuring the top state in the previous algorithm, we could measure the middle state and we will also get the desired secrete key $|\underline{\bf {\varGamma}}\rangle$

\item[$\bullet$] BVP will be solved with a single oracle, as the classical solution, if we replace the oracle $T_f$ by the oracle 
\[
\mathcal{S}_f|\underline{\bf {x}}\beta\gamma\rangle={|\underline{\bf {x}}\beta,\gamma\oplus(f(\underline{\bf {x}})\oplus\beta)\rangle}.
\]

In this case the algorithm will be the two-step algorithm:

\vskip0.1in
\noindent{\bf Step I.} Run the circuit
\begin{equation}
	\label{eq10}
	(H^{\otimes n}\otimes I\otimes I)\,\mathcal{S}_f \, (H^{\otimes n}\otimes H\,\otimes H)|\underline{\bf {0}}01\rangle.
\end{equation}

\noindent{\bf Step II.} Measure the top state, which will be certainly $|\underline{\bf \varGamma}\rangle$.
\end{enumerate}

\vskip0.1in 

The analysis can be carried out as in the previous two algorithms. The oracle  $\mathcal{S}_f$ can be also implemented to solve Deutsch-Jozsa problem.
\end{rem}

\section{Conclusions}
A quantum algorithm is proposed in terms  a CCNOT-based oracle, $T_f$, and a change of phase oracle, $P_f$, to solve BVP in a single quantum measurement. The proposed algorithm is also a polynomial speed-up algorithm. Additionally, another oracle, $S_f$, is defined to solve a new problem. Unitarity and self-adjointness of the proposed oracles are proved. This approach of adding additional qubits, or qbytes to the algorithms generally produce entanglements, which offer opportunities to solve the problems. For instance, one may consider the problem of finding two secrete strings $|\underline{\bf {\varGamma}}\rangle$ and $|\underline{\bf {\varLambda}}\rangle$, in which $f:\Bbb V\to \{0,1\}$ is a given function, $f(\underline{\bf x})=(|\underline{\bf x}\rangle+|\underline{\bf \varGamma}\rangle)\odot|\underline{\bf \varLambda}\rangle$. Furthermore, the proposed approach could be appropriately extended by using other reversible gates. However, combinations of suitable quantum oracles are accordingly required. 

\vskip0.1in

\noindent{\bf Funding.} No funds, grants, or other support was received.

\vskip0.1in 

\noindent{\bf Competing Interests.} The author have no relevant financial or non-financial interests to disclose.

  \bibliographystyle{abbrv}
  \bibliography{deutsch_refs_01}

\end{document}